\documentclass{eptcs}
\providecommand{\event}{ACT 2020} 
\usepackage{breakurl}             
\usepackage{underscore}           

\usepackage{xcolor}

\usepackage{proof}
\usepackage{stmaryrd}
\usepackage[all]{xy}
\usepackage{amssymb}
\usepackage{amsmath}
\usepackage{amsthm}
\usepackage[inline]{enumitem}
\usepackage[normalem]{ulem}
\usepackage{wasysym}

\theoremstyle{definition}

\newcommand{\I}{\mathsf{I}}
\newcommand{\ot}{\otimes}

\newcommand{\C}{\mathbb{C}}

\newcommand{\al}{\alpha}
\newcommand{\lam}{\lambda}

\newcommand{\laminv}{\lam^{-1}}
\newcommand{\alinv}{\al^{-1}}
\newcommand{\rhoinv}{\rho^{-1}}
\newcommand{\alinvc}{\al^{-1}_{\mathrm{c}}}

\newcommand{\n}{{-}}
\newcommand{\Var}{\mathsf{At}}
\newcommand{\At}{\mathsf{At}}

\newcommand{\tto}{\Longrightarrow}
\newcommand{\Tm}{\mathsf{Fma}}
\newcommand{\id}{\mathsf{id}}
\newcommand{\comp}{\circ}

\newcommand{\ax}{\mathsf{ax}}
\newcommand{\uf}{\mathsf{pass}}
\newcommand{\shift}{\uf}
\newcommand{\IL}{\I\mathsf{L}}
\newcommand{\otL}{\ot\mathsf{L}}
\newcommand{\IR}{\I\mathsf{R}}
\newcommand{\otR}{\ot\mathsf{R}}

\newcommand{\ufinv}{\mathsf{act}}

\newcommand{\scut}{\mathsf{scut}}
\newcommand{\ccut}{\mathsf{ccut}}
\newcommand{\ccutj}{\mathsf{ccut}_{\mathsf{Fma}}}
\newcommand{\ccuts}{\mathsf{ccut}_{\mathsf{Stp}}}

\newcommand{\sound}{\mathsf{sound}}
\newcommand{\cmplt}{\mathsf{cmplt}}

\newcommand{\msw}{\mathsf{switch}}
\newcommand{\mswLC}{\msw_{\mathsf{LC}}}
\newcommand{\mswRL}{\msw_{\mathsf{RL}}}

\newcommand{\switch}{\mathsf{switch}}
\newcommand{\focus}{\mathsf{focus}}
\newcommand{\emb}{\mathsf{emb}}

\newcommand{\lb}{\langle\hspace*{-0.9mm}\langle}

\newcommand{\asem}[2]{\llbracket #1 \mid #2 \rrbracket}
\newcommand{\csem}[2]{#1\, \lb #2 \rrbracket}
\newcommand{\ssem}[1]{\llbracket #1 \lb\, }
\newcommand{\morsem}[1]{\llbracket ~ \mid #1 \rrbracket} 

\newcommand{\JJ}{\mathbb{J}}

\newcommand{\Lan}{\mathrm{Lan}}

\newcommand{\Fsk}{\mathbf{Fsk}}
\newcommand{\FskRN}{\Fsk_{\mathsf{RN}}}
\newcommand{\FskLN}{\Fsk_{\mathsf{LN}}}
\newcommand{\FskAN}{\Fsk_{\mathsf{AN}}}

\newcommand\stseq[3]{#1 \mid #2 \longrightarrow #3}

\newcommand\otLcxt[1]{\ot\mathsf{C}_{#1}}
\newcommand\IC[1]{\mathsf{IC}_{#1}}
\newcommand\JJC[1]{\ot\mathsf{C}^\mathrm{c}_{#1}}
\newcommand\otC[1]{\ot\mathsf{C}_{#1}}
\newcommand\otRem{\otR_2}

\newcommand\act{\mathsf{move}}

\newcommand\ICn[1]{\mathsf{IC}^{\mathrm{c}}_{#1}}

\newcommand{\dcomp}{\mathsf{comp}}
\newcommand\stseqL[4][]{#2 \mid #3 \overset{#1}{\longrightarrow_{\mathsf{L}}} #4}
\newcommand\stseqR[4][]{#2 \mid #3 \overset{#1}{\longrightarrow_{\mathsf{R}}} #4}
\newcommand\stseqLR[4]{#1 \mid #2 \longrightarrow_{#4} #3}
\newcommand{\dottedmid}{\raisebox{-1.5pt}[0.5\height]{\hspace*{1pt}\vdots\hspace*{1pt}}}
\newcommand\stseqC[5][]{#2 \mid #3 \mathbin{\dottedmid} #4 \overset{#1}{\longrightarrow_{\mathsf{C}}} #5}

\newcommand\stseqLN[2]{#1 \longrightarrow_{\mathsf{L}} #2}
\newcommand\stseqRN[2]{#1 \longrightarrow_{\mathsf{R}} #2}

\newcommand{\highlight}[1]{\textcolor{blue}{#1}}

\newcommand{\tu}[1]{}
\newcommand{\nv}[1]{}
\newcommand{\nz}[1]{}

\title{Proof Theory of Partially Normal Skew Monoidal Categories}
\author{Tarmo Uustalu
\institute{Reykjavik University, Reykjavik, Iceland}
\institute{Tallinn University of Technology, Tallinn, Estonia}
\email{tarmo@ru.is}
\and
Niccol{\`o} Veltri
\institute{Tallinn University of Technology, Tallinn, Estonia}
\email{niccolo@cs.ioc.ee}
\and
Noam Zeilberger
\institute{{\'E}cole Polytechnique, Palaiseau, France}
\email{noam.zeilberger@lix.polytechnique.fr}
}
\def\titlerunning{Proof Theory of Partially Normal Skew Monoidal Categories}
\def\authorrunning{T. Uustalu, N. Veltri \& N. Zeilberger}
\begin{document}
\maketitle

\begin{abstract}
The skew monoidal categories of Szlach{\'a}nyi are a weakening of
monoidal categories where the three structural laws of left and right 
unitality and associativity are not required to be
isomorphisms but merely transformations in a particular direction. In
previous work, we showed that the free skew monoidal category
on a set of generating objects can be concretely
presented as a sequent calculus. This calculus enjoys cut elimination and 
admits focusing, i.e. a subsystem of canonical derivations, which solves 
the coherence problem for skew monoidal categories.

In this paper, we develop sequent calculi for partially normal skew
monoidal categories, which are skew monoidal categories with one
or more structural laws invertible. Each normality condition leads
to additional inference rules and equations on them. We prove cut
elimination and we show that the calculi admit focusing.  The result
is a family of sequent calculi between those of skew monoidal
categories and (fully normal) monoidal categories. On the level of
derivability, these define 8 weakenings of the $\I$,$\ot$ fragment
of intuitionistic non-commutative linear logic.
\end{abstract}

\section{Introduction}
\label{sec:intro}

Substructural logics are logical systems in which one or more
structural rules are not allowed. Structural rules typically include
exchange, weakening and contraction. More generally, in a sequent
calculus with sequents of the form $\Gamma \longrightarrow C$, with
the antecedent $\Gamma$ some type of a collection of formulae, a
rule is structural
if it manipulates the antecedent and
does not mention any connectives.  Affine logics are substructural
wrt.\ intuitionistic logics since contraction is disallowed. Linear
logics are substructural wrt.\ affine logics since weakening is also
disallowed. By dropping the exchange rule as well, we obtain ordered
variants of (intuitionistic) linear logics \cite{Abr:noncil}, which
include logical systems such as Lambek's syntactic calculus
\cite{Lam:matss}. One can even identify more rudimentary structural
rules, such as associativity, which is dropped in some variants of
Lambek's calculus \cite{Lam:calsty}.

Given a sequent of the form $\Gamma \longrightarrow C$ in a certain
logical system, it is natural to think of formulae in $\Gamma$ as
types of resources at our disposal, while the formula $C$ is a task
that needs to be fulfilled with the resources at hand. Under this
interpretation, the structural rules tell us how resources can be
manipulated and consumed. In intuitionistic logics, resources can be
permuted, deleted and copied.  In linear logics, they can be neither
deleted nor copied, but they can be permuted. In non-commutative
linear logics resources cannot be permuted, so they must be consumed
in the order they occur in the antecedent.

In previous work \cite{UVZ:seqcsm}, we started investigating the proof
theory of \emph{(left-)skew monoidal categories}, a weakening of
monoidal categories introduced by Szlach{\'a}nyi \cite{Szl:skemcb}
where the unitors and associator are not required to be isomorphisms
but merely transformations in a particular direction. Extending
Zeilberger's \cite{Zei:seqcsa} analysis of the Tamari order 
(which is the free (left-)skew semigroup category)
 as a sequent calculus, we introduced a sequent calculus corresponding in a
precise sense to the free skew monoidal category on a set $\At$ of
generating objects. This sequent calculus weakens the
$\I$,$\ot$-fragment of intuitionistic non-commutative linear logic
\cite{Abr:noncil} by replacing unitality and associativity with
\emph{semi-unitality} and \emph{semi-associativity}. This means that
it is possible to derive sequents corresponding to the two unitors
$\lam$ and $\rho$ and the associator $\alpha$ of skew monoidal
categories, but not sequents corresponding to the inverses of these
structural laws. Sequents have the form $\stseq S{\Gamma} C$, where
the antecedent is split into an optional formula $S$, called the
\emph{stoup}, and a list of formulae $\Gamma$, the \emph{context}. The left
rules apply only to the formula in the special stoup position. The
tensor right rule forces the formula in the stoup of the conclusion to
be the formula in the stoup of the first premise. Under the
resource-as-formulae interpretation, resources are required to be
consumed in the order they appear in the antecedent, with the formula
in the stoup being the first. At any moment, only the formula then
occupying the stoup can be decomposed on the left.

This sequent calculus enjoys cut elimination and a focused subsystem,
defining a root-first proof search strategy attempting to build a
derivation of a sequent. The focused calculus finds exactly one
representative of each equivalence class of derivations and is thus
a concrete presentation of the free skew monoidal category, as such 
solving the \emph{coherence problem} for skew monoidal categories.

Skew monoidal categories differ from normal (i.e., ordinary) monoidal
categories in that the two unitors and the associator are not
invertible. Requiring one or more of the structural laws to be
invertible, we obtain a less skew structure more like a normal
monoidal category. In this paper, we perform a proof-theoretic
analysis of each of the three normality conditions. The result is a
family of sequent calculi between those describing the free skew monoidal
category and the free monoidal category, defining altogether 8
weakenings of the $\I$,$\ot$ fragment of non-commutative intuitionistic
linear logic.

For each of these sequent calculi, we prove cut elimination and
identify a focused subsystem of canonical derivations as a concrete
presentation of the free skew monoidal category of the corresponding
degree of normality. We conclude by presenting a single parameterized focused
sequent calculus that can handle any combination of the three
normality aspects.

We fully formalized the new results presented in
Section~\ref{sec:normcond} in the dependently typed programming
language Agda. The formalization uses Agda version 2.6.0. and it is
available at
\url{https://github.com/niccoloveltri/skewmoncats-normal}.

\section{Skew Monoidal Categories}
\label{sec:skewmoncats}

A category $\C$ is said to be \emph{(left-)skew monoidal} \cite{Szl:skemcb} if it
comes together with a distinguished object $\I$, a functor
$\ot : \C \times \C \to \C$ and three natural transformations $\lam$,
$\rho$, $\al$ typed
\[
\lam_{A} : \I \ot A \to A \qquad
\rho_{A} : A \to A \ot \I \qquad
\al_{A,B,C} : (A \ot B) \ot C \to A \ot (B \ot C)
\]
satisfying the equations
\[
\small
\mathrm{(m1)}  
\xymatrix@R=1.5pc@C=0.2pc{
    & \I \ot \I \ar[dr]^-{\lambda_\I} & \\
    \I \ar[ur]^-{\rho_\I} \ar@{=}[rr] & & \I
    }
\qquad
\mathrm{(m2)}\   
\xymatrix@R=1.3pc{
      (A\ot \I) \ot B \ar[r]^{\alpha_{A,\I,B}}
      & A \ot (\I\ot B) \ar[d]^{A\ot \lambda_{B}}\\
      A \ot B \ar@{=}[r] \ar[u]^{\rho_{A}\ot B}&  A \ot B 
    }
\]
\[
\small
\mathrm{(m3)}\ 
\xymatrix@R=1.5pc@C=0.2pc{
  (\I \ot A) \ot B \ar[rr]^{\alpha_{\I,A,B}} \ar[dr] \ar@{}[d]|-{\lam_{A} \ot B}
           & &  \I \ot (A \ot B) \ar[dl]^{\lambda_{A \ot B}}\\
  & A \ot B & 
    }
\qquad
\mathrm{(m4)}\ 
\xymatrix@R=1.5pc@C=0.2pc{
  (A \ot B) \ot \I \ar[rr]^{\alpha_{A,B,\I}} 
           & &  A \ot (B \ot \I) \\
  & A \ot B \ar[ul]^{\rho_{A \ot B}} \ar[ur] & \ar@{}[u]|-{A \ot \rho_B} 
    }
\]
\[
\small
\mathrm{(m5)}\ 
\xymatrix@R=1.5pc@C=3pc{
(A\ot (B \ot C)) \ot D \ar[rr]^{\alpha_{A,B \ot C,D}}
  & & A\ot ((B \ot C)\ot D) \ar[d]^{A\ot \alpha_{B,C,D}}
  \\
((A\ot B) \ot C) \ot D \ar[u]^{\alpha_{A,B,C} \ot D}
      \ar[r]^{\alpha_{A\ot B,C,D}}
  & (A\ot B) \ot (C \ot D) \ar[r]^{\alpha_{A,B,C\ot D}}
    & A\ot (B \ot (C \ot D))
}
\]
If $\lam$, $\rho$ or $\alpha$ is a natural isomorphism, we say that
$(\C, \I, \ot)$ is \emph{left-normal}, \emph{right-normal} resp.\
\emph{associative-normal} (or \emph{Hopf}) \cite{LS:skemsw}.  
A \emph{monoidal category} \cite{Ben:catam} is a
fully normal skew monoidal category.

Equations (m1)--(m5) are directed versions of the original Mac Lane
axioms \cite{ML:natac} for monoidal categories. Kelly
\cite{Kel:maclcc} observed that, in the monoidal case, equations (m1),
(m3), and (m4) follow from (m2) and (m5). In the skew situation, this
is not the case.

Skew monoidal categories arise naturally in many settings, for example
in the study of relative monads~\cite{ACU:monnnb} and of quantum
categories~\cite{LS:skemsw}, and have been thoroughly investigated by
Street, Lack and
colleagues~\cite{LS:triosm,BGLS:catss,BL:free,BL:multi}. They,
as well as Uustalu~\cite{Uus:cohsmc}, present many examples. Here we show just one
example where normalities also play a role.

Consider two categories $\JJ$ and $\C$ and a functor $J : \JJ \to \C$
such that the left Kan extension along $J$ exists for every functor
$\JJ \to \C$.  The functor category $[\JJ, \C]$ has a skew monoidal
structure given by $\I = J$, $F \otimes G = \mathrm{Lan}_J~ F \cdot G$.
The unitors and associator are the canonical natural transformations
$\lam_F : \Lan_J~J \cdot F \to F$, $\rho_F : F \to \Lan_J~F \cdot J$,
$\al_{F,G,H} : \Lan_J~(\Lan_J~F \cdot G) \cdot H \to \Lan_J~F \cdot
\Lan_J~G \cdot H$.
This category is right-normal if $J$ is fully-faithful. It is
left-normal if $J$ is dense, which is to say that the nerve of $J$ is
fully-faithful. Finally, it is associative-normal if this nerve
preserves left Kan extensions along $J$. This example is from the work
of Altenkirch et al.~\cite{ACU:monnnb} on relative monads. Relative
monads on $J$ are (skew) monoids in the skew monoidal category
$[\JJ, \C]$.

\subsection{The Free Skew Monoidal Category}
\label{sec:freeskewmoncat}

The free skew monoidal category $\Fsk(\Var)$ over a set $\Var$ (of
atoms) can be viewed as a deductive system, which we refer to as the
skew monoidal \emph{categorical calculus}.

Objects of $\Fsk(\Var)$ are called \emph{formulae}, and are
inductively generated as follows: a formula is either an element $X$
of $\Var$ (an \emph{atom}); $\I$; or $A \ot B$ where $A$, $B$ are
formulae. We write $\Tm$ for the set of formulae.

Maps between two formulae $A$ and $C$ are \emph{derivations} of
(singleton-antecedent, singleton-succedent) sequents 
$A \tto C$, constructed using the following inference rules:
\begin{equation}\label{eq:fsk}
  \small
\renewcommand{\arraystretch}{2}
\begin{array}{c}
\infer[\id]{A \tto A}{}
\qquad
\infer[\dcomp]{A \tto C}{A \tto B & B \tto C}
\qquad
\infer[\ot]{A \ot B \tto C \ot D}{A \tto C & B \tto D}
\\
\infer[\lam]{\I \ot A \tto A}{}
\qquad
\infer[\rho]{A \tto A \ot \I}{}
\qquad
\infer[\al]{(A \ot B) \ot C \tto A \ot (B \ot C)}{}
\end{array}
\end{equation}
and \emph{identified up to} the congruence $\doteq$ induced by the equations:
\begin{equation}\label{eq:doteq}
\arraycolsep=20pt
\begin{array}{lc}
\text{(category laws)} &
\id \comp f \doteq f
\qquad
f \doteq f \comp \id
\qquad
(f \comp g) \comp h \doteq f \comp (g \comp h)
\\[6pt]
\text{($\ot$ functorial)} &
\id \ot \id \doteq \id
\qquad 
(h \comp f) \ot (k \comp g) \doteq h \ot k \comp f \ot g
\\[6pt]
&
\lam \comp \id \ot f \doteq f \comp \lam
\\
\text{($\lam,\rho,\al$ nat. trans.)} &
\rho \comp f \doteq f \ot \id \comp \rho
\\
&
\al \comp (f \ot g) \ot h \doteq f \ot (g \ot h) \comp \al
\\[6pt]
&
\lam \comp \rho \doteq \id
\qquad
\id \doteq \id \ot \lam \comp \al \comp \rho \ot \id
\\
(\textrm{m1-m5}) &
\lam \comp \al \doteq \lam \ot \id
\qquad
\al \comp \rho \doteq \id \ot \rho
\\
&
\al \comp \al \doteq \id \ot \al \comp \al \comp \al \ot \id 
\end{array}
\end{equation}
In the term notation for derivations, we write $g \comp f$ for
$\dcomp\, f\, g$ to agree with the standard categorical notation.

Mac Lane's coherence theorem \cite{ML:natac} says that, in the free monoidal
category, there is exactly one map $A \tto B$ if the formulae $A$ and
$B$ have the same frontier of atoms and no such map otherwise.

For the free skew monoidal category, this does not hold. We have pairs
of formulae that have the same frontier of atoms but no maps between
them or multiple maps. There are no maps
$X \tto \I \ot X$, no maps $X \ot \I \tto X$ and no maps
$X \ot (Y \ot Z) \tto (X \ot Y) \ot Z$.  At the same time, we have
two maps $\id \not\doteq \alpha \comp \rho \ot \lambda : X \ot (\I \ot Y) \tto X
\ot (\I \ot Y)$
and two maps
$\id \not\doteq \rho \ot \lambda \comp \alpha : (X \ot \I) \ot Y \tto (X
\ot \I) \ot Y$.

\subsection{Skew Monoidal Sequent Calculus}
\label{sec:seqcalcskewmon}
  
In \cite{UVZ:seqcsm}, we showed that the free skew monoidal category
$\Fsk(\At)$ admits an equivalent presentation as a sequent
calculus. In the latter, sequents are triples $\stseq S {\Gamma}
C$. The antecedent is a pair of a \emph{stoup} $S$ together with a
\emph{context} $\Gamma$, while the succedent $C$ is a single formula.
A stoup is an optional formula, meaning that it can either be empty
(written $\n$) or contain a single formula. A context is a list of
formulae. For the empty list, we usually just leave a space, 
but where necessary for readability, we write $()$. 
Derivations in the sequent calculus are inductively
generated by these inference rules:
\begin{equation}\label{eq:seqcalc}
  \small
\begin{array}{c@{\quad \quad}c@{\quad \quad}c}
\infer[\uf]{\stseq{\n} {A, \Gamma}{C}}{
  \stseq{A}{\Gamma} C
}
&
\infer[\IL]{\stseq{\I} {\Gamma}{C}}{
  \stseq{\n}{\Gamma}{C}
}
&
\infer[\otL]{\stseq{A \ot B}{\Gamma} C}{
  \stseq{A}{B, \Gamma} C
}
\\[6pt]
\infer[\ax]{\stseq A {~} A}{
}
&
\infer[\IR]{\stseq{\n}{~}{\I}}{
}
&
\infer[\otR]{\stseq{S}{\Gamma, \Delta}{A \otimes B}}{
  \stseq{S}{\Gamma}{A}
  &
  \stseq{\n}{\Delta} B
}
\end{array}
\end{equation}
($\uf$ for `passivate', $\mathsf{L}$, $\mathsf{R}$ for introduction
on the left (in the stoup) resp.\ right)
and identified up to the congruence $\circeq$ induced by the equations:
\begin{equation}\label{eq:circeq}
\small
\begin{array}{@{\qquad}c@{\qquad}l}
  \multicolumn{2}{l}{\textrm{($\eta$-conversions)}} \\[6pt]
  \multicolumn{2}{c}
  {\ax_{\I} \circeq \IL\;\IR
  \qquad\qquad
  \ax_{A \ot B} \circeq \otL\;(\otR\;(\ax_A,\uf\;\ax_B))}
  \\[12pt]
    \multicolumn{2}{l}{\textrm{(commutative conversions)}} \\[6pt]
  \otR\;(\uf\;f,g) \circeq \uf\;(\otR\;(f,g))
  & (\text{for } f : \stseq{A'}{\Gamma}{A}, \; g : \stseq{\n}{\Delta}B)
  \\[6pt]
  \otR\;(\IL\;f,g) \circeq \IL\;(\otR\;(f,g))
  & (\text{for } f : \stseq{\n}{\Gamma}{A}, \; g : \stseq{\n}{\Delta}B)
  \\[6pt]
  \otR\;(\otL\;f,g) \circeq \otL\;(\otR\;(f,g))
  & (\text{for } f : \stseq{A'}{B',\Gamma}{A}, \; g : \stseq{\n}{\Delta}B)
\end{array}
\end{equation}

Although these rules look very similar to the rules of the $\I$, $\ot$
fragment of intuitionistic non-commuta\-tive linear logic
\cite{Abr:noncil} (the sequent calculus describing the free monoidal category)---in
particular, there is no left exchange rule, weakening or
contraction---, there are two crucial differences.
\begin{itemize}
\item The left rules $\IL$ and $\otL$ are restricted to apply only to
  the formula within the stoup (in the conclusion-first reading of these rules).
  This restriction was also present in
  Zeilberger's sequent calculus for the Tamari order
  \cite{Zei:seqcsa}. In this calculus, it is possible to derive sequents
  corresponding to the right unitor $\rho : A \tto A \ot \I$ and the
  associator $\alpha : (A \ot B) \ot C \tto A \ot (B \ot C)$:
  \[
  \small
  \infer[\otR]{\stseq A {~} {A \ot \I}}{
    \infer[\ax]{\stseq{A}{~}{A}}{}
    &
    \infer[\IR]{\stseq {\n}{~} \I}{}
  }
  \qquad
  \infer[\otL]{\stseq {(A \ot B) \ot C}{~}{A \ot (B \ot C)}}{
    \infer[\otL]{\stseq{A\ot B}{C}{A\ot (B \ot C)}}{
      \infer[\otR]{\stseq A {B,C}{A \ot (B \ot C)}}{
        \infer[\ax]{\stseq A {~} A}{}
        &
        \infer[\uf]{\stseq{\n}{B,C}{B \ot C}}{
          \infer[\otR]{\stseq{B}{C}{B\ot C}}{
            \infer[\ax]{\stseq B {~} B}{}
            &
            \infer[\uf]{\stseq {\n} {C} C}{
              \infer[\ax]{\stseq C {~}C}{}
            }
          }
        }
      }
    }
  }
  \]
  On the other hand, since $\IL$ and $\otL$ only act on the formula in
  the stoup, there is no way of deriving sequents corresponding to
  inverses of $\rho$ and $\alpha$ for atomic $A$ resp.\ $A$, $B$, $C$.
  
\item The stoup is allowed to be empty, permitting a distinction
  between antecedents of the form $A \mid \Gamma$ (with $A$ inside the
  stoup) and antecedents of the form $\n \mid A, \Gamma$ (with $A$
  outside the stoup). This distinction plays an important role in the
  right rule $\otR$, which sends the formula in the stoup, when it is
  present, to the first premise. In this calculus, it is possible to
  derive a sequent corresponding to the left unitor $\lam : \I \ot A
  \tto A$:
  \[
  \small
  \infer[\otL]{\stseq {\I \ot A}{~}{A}}{
    \infer[\IL]{\stseq {\I} {A}{A}}{
      \infer[\uf]{\stseq{\n}{A}{A}}{
        \infer[\ax]{\stseq A {~}{A}}{}
      }
    }
  }  
  \]
  On the other hand, since $\otR$ always send the formula in the stoup
  to the first premise, it is not possible to derive a sequent
  corresponding to the inverse of $\lam$ for atomic $A$.
\end{itemize}

There are no primitive cut rules in this sequent calculus, but two forms of
cut are admissible:
\begin{equation}\label{eq:cut}
  \small
\infer[\scut]{\stseq S {\Gamma, \Delta} C}{
  \stseq S {\Gamma} A
  &
  \stseq{A} {\Delta} C
}
\qquad
\infer[\ccut]{\stseq S {\Delta_0, \Gamma, \Delta_1} C}{
  \stseq {\n} {\Gamma} A
  &
  \stseq {S} {\Delta_0, A, \Delta_1} C
}
\end{equation}

Sequent calculus derivations can be turned into categorical calculus
derivations by means of a function $\sound : \stseq S {\Gamma} C \to
\asem{S}{\Gamma} \tto C$, where the interpretation of an
antecedent as a formula $\asem{S}{\Gamma}$ is defined as $\asem{S}{\Gamma} = \csem{\ssem{S}}{\Gamma}$ with
\[
\ssem{\n} = \I \qquad \ssem{A} = A \qquad\qquad \csem{A}{~} = A \qquad
\csem{A}{B,\Gamma} = \csem{(A \ot B)}{\Gamma}
\]
which means that $\csem{A}{A_1, A_2 \ldots, A_n} = (\ldots (A \otimes
A_1) \otimes A_2) \ldots) \otimes A_n$.
The interpretation of antecedents is functorial, i.e., we 
have the following inference rule:
\begin{equation}\label{eq:morsem}
  \small
  \infer[\morsem{\Gamma}]{\asem{A}{\Gamma} \tto \asem{B}{\Gamma}}{
    A \tto B
  }
\end{equation}
The function $\sound$ is well-defined on $\circeq$-equivalence
classes: given related derivations $f \circeq g$, then $\sound\;f
\doteq \sound \;g$.

Categorical calculus derivations can be interpreted as sequent
calculus derivations via a function $\cmplt : \asem{S}{\Gamma}
\tto C \to \stseq S {\Gamma} C$, well-defined on $\doteq$-equivalence
classes: given related derivations $f \doteq g$, then $\cmplt\;f
\circeq \cmplt \;g$. The cut rule $\scut$ is fundamental for modelling
the composition operation $\dcomp$ of the categorical calculus.

The functions $\sound$ and $\cmplt$ establish a bijection between
derivations in the categorical calculus and the sequent calculus:
$\sound\;(\cmplt\;f) \doteq f$ and $\cmplt\;(\sound\;g) \circeq g$,
for all $f : \asem{S}{\Gamma} \tto C$ and $g : \stseq S {\Gamma} C$.

\subsection{A Focused Subsystem}
\label{sec:focusskewmon}

The congruence relation $\circeq$ can be considered as a term rewrite
system, by directing every equation from left to right. The resulting
rewrite system is weakly confluent and strongly normalizing, hence
confluent with unique normal forms.

Normal-form derivations in our sequent calculus can be described as
derivations in a \emph{focused} subsystem.
 In the style of Andreoli \cite{And:logpfp}, we present the
focused subsystem as a sequent calculus with an additional \emph{phase
  annotation} on sequents. In phase $\mathsf{L}$, sequents are of
the form $\stseqL S {\Gamma} C$, where $S$ is a general stoup. In 
phase $\mathsf{R}$, sequents take the form $\stseqR T {\Gamma} C$,
where $T$ is an \emph{irreducible} stoup, that is an optional atom:
either empty or an atomic formula.  Derivations in the focused 
calculus are inductively generated by the following inference rules:
\begin{equation}\label{eq:focusseqcalc}
  \small
  \begin{array}{c}
      \infer[\uf]{\stseqL{\n}{A, \Gamma} C}{
        \stseqL A {\Gamma} C
      }
      \qquad
      \infer[\IL]{\stseqL{\I}{\Gamma}{C}}{
        \stseqL{\n}{\Gamma} C
      }
      \qquad
      \infer[\otL]{\stseqL{A \ot B}{\Gamma} C}{
        \stseqL A {B, \Gamma} C
      }
      \qquad
      \infer[\msw]{\stseqL T {\Gamma} C}{
        \stseqR T {\Gamma} C
      }
\\[6pt]    
    \infer[\ax]{\stseqR X {~} X}{
    }
    \qquad
    \infer[\IR]{\stseqR{\n}{~} {\I}}{
    }
    \qquad
    \infer[\otR]{\stseqR T {\Gamma, \Delta} {A \otimes B}}{
      \stseqR T {\Gamma} A
      &
      \stseqL{\n} {\Delta} B
    } 
  \end{array}
\end{equation}

The focused rules define a sound and complete proof search strategy.
The focused calculus is clearly sound: by erasing phase annotations,
all of the rules are either rules of the original calculus or else (in
the case of $\switch$) have the conclusion equal to the premise. So
focused calculus derivations can be embedded into sequent calculus
derivations via a function $\emb_P: \stseqLR{S}{\Gamma}{C}{P} \to
\stseq S \Gamma C$, where $P \in \{\mathsf{L}, \mathsf{R}\}$.

The focused calculus is also complete: we can define a normalization
procedure $\focus : \stseq S \Gamma C \to \linebreak \stseqL{S}{\Gamma}{C}$
sending each derivation in the sequent calculus to a
canonical representative of its $\circeq$-equivalence class in the
focused calculus. This means in particular that $\focus$ maps
$\circeq$-related derivations to \emph{equal} focused derivations.

The functions $\focus$ and $\emb_{\mathsf{L}}$ establish a bijection
between derivations in the full sequent calculus (up to $\circeq$) and
its focused subsystem: $\emb_{\mathsf{L}}\; (\focus\;f) \circeq f$ and
$\focus\;(\emb_{\mathsf{L}}\;g) = g$, for all
$f : \stseq S {\Gamma} C$ and $g : \stseqL S {\Gamma} C$.

Putting this together with the results discussed in
Section~\ref{sec:seqcalcskewmon}, it follows that the focused 
calculus is a concrete presentation of the free skew monoidal category
$\Fsk(\At)$. As such, the focused calculus solves the problem
of characterizing the homsets of the free skew monoidal category,
a.k.a.\ the \emph{coherence problem}. Moreover, it solves
two related algorithmic problems effectively:
\begin{itemize}
\item Duplicate-free enumeration of all maps $A \tto C$ in the form of
  representatives of $\doteq$-equivalence classes of categorical
  calculus derivations: For this, find all focused derivations of
  $\stseqL A {~} C$, which is solvable by exhaustive proof search, which
  terminates, and
  translate them to the categorical calculus derivations.
\item Finding whether two given maps of type $A \tto C$, presented as
  categorical calculus derivations, are equal, i.e.,
  $\doteq$-related as derivations: For this, translate them to focused
  derivations of $\stseqL A {~} C$ and check whether they are
  equal, which is decidable.
\end{itemize}

Different approaches to the coherence problem of skew monoidal
categories have been considered before. Uustalu \cite{Uus:cohsmc}
identified a class of normal forms of objects in $\Fsk{(\At)}$ and
showed that there exists at most one map between an object and an
object in normal form, and exactly one map between an object and
that object's normal form. In another direction, Lack and Street
\cite{LS:triosm} addressed the problem of determining equality of maps
by proving that there is a faithful, structure-preserving functor
$\Fsk{(1)} \to \Delta_\bot$ from the free skew monoidal category on one
generating object to the category of finite non-empty ordinals and
first-element-and-order-preserving functions (which is an
associative-normal skew-monoidal category under the ordinal sum with
unit 1).  This approach was further elaborated by Bourke and Lack
\cite{BL:free} with a more explicit description of the homsets of
$\Fsk{(1)}$.

Let us use the focused calculus to analyze where multiple maps
$A \tto C$ can come from. There are two sources of non-determinism in
root-first proof search in the focused calculus: (i) in phase $\mathsf{L}$,
when the stoup is empty, whether to apply $\uf$ or $\msw$, and (ii) in
phase $\mathsf{R}$, when the succedent formula is a tensor and the
rule $\otR$ is to be applied, how to split the context into $\Gamma$
and $\Delta$. In the latter situation, only those choices where
$|T|, |\Gamma| = |A|$ and $|\Delta| = |B|$ can
possibly lead to a complete derivation;
we write $|~|$ for the frontier of atoms in a
formula, an optional formula or a list of formulae. But there can be
multiple such choices if in the middle of the context there are closed
formulae (i.e., formulae made of $\I$ and $\ot$ only): 
those can be freely split between $\Gamma$ and $\Delta$.

The two maps
$\id \not\doteq \alpha \comp \rho \ot \lambda : X \ot (\I \ot Y) \tto X
\ot (\I \ot Y)$
translate to two different focused derivations of the sequent
$\stseqL{X \ot (\I \ot Y)}{~}{X \ot (\I \ot Y)}$ because of
the non-determinism of type (i). Notice the choice
between $\uf$ and $\msw$.
\begin{equation}\label{eq:nondet1}
\scriptsize
\begin{array}{c}
\infer[\otL]{\stseqL {X \ot (\I \ot Y)} {~} {X \ot (\I \ot Y)}}{
  \infer[\msw]{\stseqL {X} {\I \ot Y} {X \ot (\I \ot Y)}}{
    \infer[\otR]{\stseqR {X} {\I \ot Y} {X \ot (\I \ot Y)}}{
      \infer[\ax]{\stseqR {X} {~} {X}}{}
      &
      \infer[\highlight{\uf}]{\stseqL{\n}{\I \ot Y}{\I \ot Y}}{
        \infer[\otL]{\stseqL{\I \ot Y}{~}{\I \ot Y}}{
          \infer[\IL]{\stseqL{\I}{Y}{\I \ot Y}}{
            \infer[\msw]{\stseqL{\n}{Y}{\I \ot Y}}{
              \infer[\otR]{\stseqR{\n}{Y}{\I \ot Y}}{
                \infer[\IR]{\stseqR{\n}{~}{\I}}{}
                &
                \infer[\uf]{\stseqL{\n}{Y}{Y}}{
                  \infer[\msw]{\stseqL{Y}{~}{Y}}{
                    \infer[\ax]{\stseqR{Y}{~}{Y}}{}
                  }
                }
              }
            }
          }
        }
      }
    }
  }
}
\qquad
\infer[\otL]{\stseqL {X \ot (\I \ot Y)} {~} {X \ot (\I \ot Y)}}{
  \infer[\msw]{\stseqL {X} {\I \ot Y} {X \ot (\I \ot Y)}}{
    \infer[\otR]{\stseqR {X} {\I \ot Y} {X \ot (\I \ot Y)}}{
      \infer[\ax]{\stseqR {X} {~} {X}}{}
      &
      \infer[\highlight{\msw}]{\stseqL{\n}{\I \ot Y} {\I \ot Y}}{
        \infer[\otR]{\stseqR{\n}{\I \ot Y} {\I \ot Y}}{
          \infer[\IR]{\stseqR{\n}{~} {\I}}{}
          &
          \infer[\uf]{\stseqL{\n}{\I \ot Y} {Y}}{
            \infer[\otL]{\stseqL{\I \ot Y}{~} {Y}}{
              \infer[\IL]{\stseqL{\I}{Y} {Y}}{
                \infer[\uf]{\stseqL{\n}{Y} {Y}}{
                  \infer[\msw]{\stseqL{Y}{~} {Y}}{
                    \infer[\ax]{\stseqR{Y}{~} {Y}}{}
                  }
                }
              }
            }
          }
        }
      }
    }
  }
}
\end{array}
\end{equation}

The two maps
$\id \not\doteq \rho \ot \lambda \comp \alpha : (X \ot \I) \ot Y \tto (X
\ot \I) \ot Y$
translate to distinct focused derivations of the sequent
$\stseqL{(X \ot \I) \ot Y}{~}{(X \ot \I) \ot Y}$ due to type-(ii) non-determinism. Here the first (from the endsequent) application of the tensor
right rule $\otR$ splits the context in two different ways: in the
first derivation the unit in the context is sent to the first
premise, while in the second derivation it is sent to the second
premise.
\begin{equation}\label{eq:nondet2}
\scriptsize
\begin{array}{c}
\infer[\otL]{\stseqL {(X \ot \I) \ot Y}{~} {(X \ot \I) \ot Y}}{
  \infer[\otL]{\stseqL {X \ot \I} {Y} {(X \ot \I) \ot Y}}{
    \infer[\msw]{\stseqL {X}{\I,Y} {(X \ot \I) \ot Y}}{
      \infer[\highlight{\otR}]{\stseqR {X}{\I,Y} {(X \ot \I) \ot Y}}{
        \infer[\otR]{\stseqR {X}{\highlight{\I}}{X \ot \I}}{
          \infer[\ax]{\stseqR {X}{~}{X}}{}
          &
          \infer[\uf]{\stseqL {\n}{\I}{\I}}{
            \infer[\IL]{\stseqL {\I}{~}{\I}}{
              \infer[\msw]{\stseqL {\n}{~}{\I}}{
                \infer[\IR]{\stseqR {\n}{~}{\I}}{}
              }
            }
          }
        }
        &
        \infer[\uf]{\stseqL {\n}{Y} {Y}}{
          \infer[\msw]{\stseqL {Y}{~} {Y}}{
            \infer[\ax]{\stseqR {Y}{~} {Y}}{}
          }
        }
      }
    }
  }
}
\qquad 
\infer[\otL]{\stseqL {(X \ot \I) \ot Y}{~} {(X \ot \I) \ot Y}}{
  \infer[\otL]{\stseqL {X \ot \I} {Y} {(X \ot \I) \ot Y}}{
    \infer[\msw]{\stseqL {X}{\I,Y} {(X \ot \I) \ot Y}}{
      \infer[\highlight{\otR}]{\stseqR {X}{\I,Y} {(X \ot \I) \ot Y}}{
        \infer[\otR]{\stseqR {X}{~} {X \ot \I}}{
          \infer[\ax]{\stseqR {X}{~} {X}}{}
          &
          \infer[\msw]{\stseqL {\n}{~} {\I}}{
            \infer[\IR]{\stseqR {\n}{~} {\I}}{}
          }
        }
        &
        \infer[\uf]{\stseqL {\n}{\highlight{\I},Y} {Y}}{
          \infer[\IL]{\stseqL {\I}{Y} {Y}}{
            \infer[\uf]{\stseqL {\n}{Y} {Y}}{
              \infer[\msw]{\stseqL {Y}{~} {Y}}{
                \infer[\ax]{\stseqR {Y}{~} {Y}}{}
              }
            }
          }
        }
      }
    }
  }
}
\end{array}
\end{equation}

\section{Normality Conditions}
\label{sec:normcond}

\subsection{Left-Normality}
\label{sec:left}


The free left-normal skew monoidal category $\FskLN(\At)$ on a set
$\At$ is obtained by extending the grammar of derivations in the
fully skew categorical calculus (\ref{eq:fsk}) with a new inference rule:
\[
\small
\infer[\laminv]{A \tto \I \ot A}{}
\]
and extending the equivalence of derivations (\ref{eq:doteq}) with two
new equations:
$\lam \comp \laminv \doteq \id$ and $\laminv \comp \lam \doteq \id$.

An equivalent sequent calculus presentation of $\FskLN(\At)$ is
obtained by adding another right rule for the tensor $\ot$ to the fully skew
sequent calculus (\ref{eq:seqcalc}), which allows one to send the
formula in the stoup to the second premise, provided that all of the context 
is also sent to the second premise (so the
antecedent of the first premise is left completely empty).
\[
\small
\infer[\otRem]{\stseq{A'} {\Delta}{A \ot B}}{
  \stseq{\n}{~}{A} & \stseq{A'}{\Delta}{B}
}
\]
The introduction of $\otRem$ makes it possible to derive a sequent
corresponding to $\laminv : A \tto \I \ot A$:
\[
\small
\infer[\otRem]{\stseq A {~}{\I \ot A}}{
  \infer[\IR]{\stseq {\n}{~}{\I}}{}
  &
  \infer[\ax]{\stseq{A}{~}{A}}{}
  }
\]
In particular this allows us to interpret categorical derivations into
sequent calculus derivations for the left-normal case, extending the
definition of the function $\cmplt$ introduced in
Section~\ref{sec:seqcalcskewmon}.

Equivalence of derivations in the sequent calculus is the least
congruence $\circeq$ induced by the equations in (\ref{eq:circeq})
together with the following equations:
\[
\small
\begin{array}{c@{\qquad}l}
  \otR\;(f,\uf\;g) \circeq \uf\;(\otRem\;(f,g))
  & (\text{for } f : \stseq{\n}{~}{A}, \; g : \stseq{A'}{\Delta}{B})
  \\[6pt]
  \otRem\;(f,\IL\;g) \circeq \IL\;(\otR\;(f,g))
  & (\text{for } f : \stseq{\n}{~}{A}, \; g : \stseq{\n}{\Delta}{B})
  \\[6pt]
  \otRem\;(f,\otL\;g) \circeq \otL\;(\otRem\;(f,g))
  & (\text{for } f : \stseq{\n}{~}{A}, \; g : \stseq{A'}{B',\Delta}{B})
\end{array}
\]

The two cut rules in (\ref{eq:cut}) are also admissible in the
left-normal sequent calculus. The definition of $\ccut$ uses the following
rule $\ufinv$ (for `activate'), admissible in this sequent calculus (but not in the fully skew one) and inverting $\uf$ up to the
congruence $\circeq$:
\[
\small
\infer[\ufinv]{\stseqL A {\Gamma} C}{
  \stseqL{\n}{A, \Gamma} C        
}
\]

The transformation $\sound$ introduced in
Section~\ref{sec:seqcalcskewmon}, interpreting the fully skew sequent calculus
derivations as categorical calculus derivations, extends to the
left-normal case. Given $f : \stseq{\n}{~}{A}$ and $g :
\stseq{A'}{\Delta}{B}$, define $\sound\;(\otRem(f,g))$ as:
\[
\small
\infer[\dcomp]{\asem {A'} {\Delta} \tto A \ot B}{
  \asem {A'} {\Delta} \overset{\sound\,g}{\tto} B
  &
  \infer[\dcomp]{B \tto A \ot B}{
    \infer[\laminv]{B \tto \I \ot B}{}
    &
    \I \ot B \overset{\sound\,f \ot \id}{\tto} A \ot B
  }
}
\]

Again, the congruence relation $\circeq$ read as a term rewrite system
is weakly confluent and strongly normalizing, and normal-form
derivations in the sequent calculus can be described as derivations in
a focused subsystem. Sequents in the left-normal focused calculus are
annotated with two possible phase annotations, as in the fully skew
focused calculus (\ref{eq:focusseqcalc}). The condition for
switching phase is different: the formula in the stoup is still
required to be irreducible, but if the stoup is empty we are allowed
to switch phase only when the context is empty as well. In phase
$\mathsf{R}$, we also include the new tensor right rule $\otRem$, in
which both premises are required to be $\mathsf{R}$-phase derivations.
Again $T$ is an irreducible stoup: either empty or an atomic formula.
\[
\small
\begin{array}{c}
  \infer[\uf]{\stseqL{\n}{A, \Gamma} C}{
    \stseqL A {\Gamma} C
  }
  \quad
  \infer[\IL]{\stseqL{\I}{\Gamma}{C}}{
    \stseqL{\n}{\Gamma} C
  }
  \quad
  \infer[\otL]{\stseqL{A \ot B}{\Gamma} C}{
    \stseqL A {B,\Gamma} C
  }
  \quad 
  \infer[\msw]{\stseqL T {\Gamma} C}{
    \stseqR T {\Gamma} C
    &
    T = \n \to \Gamma = ()
  }
  \\[6pt]
  \infer[\ax]{\stseqR X {~} X}{
  }  
  \quad
  \infer[\IR]{\stseqR{\n}{~} {\I}}{
  }
  \quad 
  \infer[\otR]{\stseqR T {\Gamma, \Delta} {A \otimes B}}{
    \stseqR T {\Gamma} A
    &
    \stseqL{\n} {\Delta} B
  }
  \quad
  \infer[\otRem]{\stseqR {X} {\Delta} {A \otimes B}}{
    \stseqR {\n}{~} A
    &
    \stseqR{X} {\Delta} B
  }
\end{array}
\]

All $\mathsf{R}$-phase sequents in a derivation of a sequent
$\stseqL S \Gamma C$ have the context empty if the stoup is empty.

The functions $\emb_{\mathsf{L}}$ and $\emb_{\mathsf{R}}$ embedding
fully skew focused calculus derivations in the unfocused sequent
calculus, discussed in Section~\ref{sec:focusskewmon}, can clearly be
adapted to the left-normal case. The same holds for the $\focus$ function,
sending a sequent calculus derivation to its
$\circeq$-normal form in the focused calculus. In particular,
this means that $\focus$ maps two $\circeq$-equivalent derivations 
to the same focused
derivation. Similarly to the fully skew case,
$\focus$ and $\emb_\mathsf{L}$ establish a bijection between maps in the
left-normal sequent calculus (up to $\circeq$) and its focused subsystem.

The left-normal focused calculus has less non-determinism than the
fully skew focused calculus. The non-determinism of type (i) is not
there: In phase $\mathsf{L}$, $\msw$ cannot be applied unless the
context is empty while $\uf$ only applies if it is non-empty. The
non-determinism of type (ii) remains much like in the fully skew case
except that there are two tensor right-rules, $\otR$ and $\otRem$.  In
a choice that can lead to a completed derivation, we must be able 
to take $|T|, |\Gamma| = |A|$ and $|\Delta| = |B|$ in $\otR$ or
$() = |A|$ and $X, |\Delta| = |B|$ in $\otRem$. There may be multiple
such choices if in the middle of the context we have closed formulae. 

In the free left-normal skew monoidal category, we have
$\id \doteq \alpha \comp \rho \ot \lambda : X \ot (\I \ot
Y) \tto X \ot (\I \ot Y)$:
\[
\id
\doteq \id \ot \laminv \comp \id \ot \lam
\doteq \id \ot \laminv \comp (\id \ot \lam \comp \al \comp \rho \ot \id) \comp \id \ot \lam
\doteq \alpha \comp \rho \ot \lambda
\]
This collapse is reflected by there being exactly one focused
derivation of the sequent
$\stseqL{X \ot (\I \ot Y)}{~}{X \ot (\I \ot Y)}$.
Compare this with the two distinct derivations of the sequent in the
fully skew focused calculus, displayed in~(\ref{eq:nondet1}). In the left-normal focused calculus, we are forced to use $\uf$, we cannot apply $\msw$ since the stoup is empty but the context is not.
\[
\small 
\infer[\otL]{\stseqL {X \ot (\I \ot Y)} {~} {X \ot (\I \ot Y)}}{
  \infer[\msw]{\stseqL {X} {\I \ot Y} {X \ot (\I \ot Y)}}{
    \infer[\otR]{\stseqR {X} {\I \ot Y} {X \ot (\I \ot Y)}}{
      \infer[\ax]{\stseqR {X} {~} {X}}{}
      &
      \infer[\highlight{\uf}]{\stseqL{\n}{\I \ot Y}{\I \ot Y}}{
        \infer[\otL]{\stseqL{\I \ot Y}{~}{\I \ot Y}}{
          \infer[\IL]{\stseqL{\I}{Y}{\I \ot Y}}{
            \infer[\uf]{\stseqL{\n}{Y}{\I \ot Y}}{
              \infer[\msw]{\stseqL{Y}{~}{\I \ot Y}}{
                \infer[\otRem]{\stseqR{Y}{~}{\I \ot Y}}{
                  \infer[\IR]{\stseqR{\n}{~}{\I}}{}
                  &
                  \infer[\ax]{\stseqR{Y}{~}{Y}}{}                  
                }
              }
            }
          }
        }
      }
    }
  }
}
\]

\smallskip

The left-normal sequent calculus admits
also a stoup-free presentation. This is because the rule $\uf$
is invertible. Here we only show the focused subsystem of the stoup-free
variant. In phase $\mathsf{L}$, sequents are of the form $\stseqLN
{\Gamma} C$, where $\Gamma$ is a general list of formulae. In
phase $\mathsf{R}$, sequents take the form $\stseqRN {\Lambda} C$,
where $\Lambda$ is an irreducible list of formulae, meaning that it is
either empty or the formula in its head is an atom.
\[
\small
\begin{array}{c}
  \infer[\IL]{\stseqLN{\I,\Gamma}{C}}{
    \stseqLN{\Gamma} C
  }
  \qquad
  \infer[\otL]{\stseqLN{A \ot B,\Gamma} C}{
    \stseqLN{A,B,\Gamma} C
  }
  \qquad
  \infer[\msw]{\stseqLN {\Lambda} C}{
    \stseqRN {\Lambda} C
  }
  \\[6pt]
  \infer[\ax]{\stseqRN X X}{
  }  
  \qquad
  \infer[\IR]{\stseqRN{~} {\I}}{
  }
  \qquad
  \infer[\otR]{\stseqRN {\Lambda, \Delta} {A \otimes B}}{
    \stseqRN {\Lambda} A
    &
    \stseqLN {\Delta} B
  }
\end{array}
\]
Derivations of a sequent
$\stseqL S {\Gamma} C$ are in a bijection with derivations of
$\stseqLN {\ssem{S},\Gamma} C$ where $\ssem{S}$ is the interpretation
of a stoup as a formula introduced in
Section~\ref{sec:seqcalcskewmon}.

\subsection{Right-Normality}
\label{sec:right}


The free right-normal skew monoidal category $\FskRN(\At)$ on a set
$\At$ is obtained by extending the grammar of derivations in the
fully skew categorical calculus (\ref{eq:fsk}) with a new inference rule:
\[
\small
\infer[\rhoinv]{A \ot \I \tto A}{}
\]
and extending the equivalence of derivations (\ref{eq:doteq}) with two
new equations: $\rho \comp \rhoinv \doteq \id$ and $\rhoinv \comp \rho
\doteq \id$.

An equivalent sequent calculus presentation of $\FskRN(\At)$ is
realized by adding additional left rules for $\I$ and $\ot$ to the
fully skew sequent calculus (\ref{eq:seqcalc}), relaxing the
condition for deleting the unit and decomposing tensors in the antecedent:
\[
\small
\infer[\IC{\Gamma_0}]{\stseq{S} {\Gamma_0,\I,\Gamma_1}{C}}{
  \stseq{S}{\Gamma_0,\Gamma_1}{C}
}
\qquad
\infer[\JJC{\Gamma_0}]{\stseq{S} {\Gamma_0,J \ot J',\Gamma_1}{C}}{
  \stseq{S}{\Gamma_0,J,J',\Gamma_1}{C}
}
\]
Here and later, $J$ and $J'$ stand for closed formulae.
The introduction of $\IC{}$ makes it possible to derive a
sequent corresponding to $\rhoinv : A \ot \I \tto A$ for any $A$,
including $A = X$:
\[
\small
\infer[\otL]{\stseq{A \ot \I}{~}{A}}{
  \infer[\IC{()}]{\stseq A {\I}{A}}{
    \infer[\ax]{\stseq A{~} A}{}
  }
}
\]

The rule $\JJC{}$ is needed since it is important to allow deletion in
the context of any closed formula and not just $\I$: we need to be
able to derive, e.g., the sequent $\stseq{X}{\I \ot
  \I}{X}$. Analogously, in the left-normal sequent calculus of
Section~\ref{sec:left}, it was important to be able to derive the
sequent $\stseq{X}{~}{(\I \ot \I) \ot X}$, which was possible since the
first premise of the inference rule $\otRem$ is a sequent
$\stseq{\n}{~}{A}$, which is derivable precisely when $A$ is closed.

Equivalence of derivations in the sequent calculus is the least
congruence $\circeq$ induced by the equations in (\ref{eq:circeq})
together with the following equations:

\begin{equation}\label{eq:circeq-rn}
\small
\begin{array}{c@{\qquad}l}
  \uf\;(\IL\;f) \circeq \IC{()}\;f
  & (\text{for } f : \stseq{\n}{\Gamma}{C})
  \\[12pt]
  \IC{\Gamma_0}\;(\IC{\Gamma_0,\Gamma_1}\;f) \circeq \IC{\Gamma_0,\I,\Gamma_1}\;(\IC{\Gamma_0}\;f)
  & (\text{for } f : \stseq{S}{\Gamma_0,\Gamma_1,\Gamma_2}{C})
  \\[6pt]
  \JJC{\Gamma_0}\;(\IC{\Gamma_0,J,J',\Gamma_1}\;f) \circeq \IC{\Gamma_0,J\ot J',\Gamma_1}\;(\JJC{\Gamma_0}\;f)
  & (\text{for } f : \stseq{S}{\Gamma_0,J,J',\Gamma_1,\Gamma_2}{C})
  \\[6pt]
  \uf\;(\IC{\Gamma_0}\;f) \circeq \IC{A,\Gamma_0}\;(\uf\;f)
  & (\text{for } f : \stseq{A}{\Gamma_0,\Gamma_1}{C})
  \\[6pt]
  \IL\;(\IC{\Gamma_0}\;f) \circeq \IC{\Gamma_0}\;(\IL\;f)
  & (\text{for } f : \stseq{\n}{\Gamma_0,\Gamma_1}{C})
  \\[6pt]
  \otL\;(\IC{B,\Gamma_0}\;f) \circeq \IC{\Gamma_0}\;(\otL\;f)
  & (\text{for } f : \stseq{A}{B,\Gamma_0,\Gamma_1}{C})
  \\[6pt]
  \otR\;(\IC{\Gamma_0}\;f,g) \circeq \IC{\Gamma_0}\;(\otR\;(f,g))
  & (\text{for } f : \stseq{S}{\Gamma_0,\Gamma_1}{A}, \; g : \stseq{\n}{\Delta}B)
  \\[6pt]
  \otR\;(f,\IC{\Delta_0}\;g) \circeq \IC{\Gamma,\Delta_0}\;(\otR\;(f,g))
  & (\text{for } f : \stseq{S}{\Gamma}{A}, \; g : \stseq{\n}{\Delta_0,\Delta_1}B)
\end{array}
\end{equation}
plus the same number of similar equations for $\JJC{}$.
The
only difference is in the 1st equation, with $\otL$ instead of $\IL$ in the left-hand side, in which an extra
application of $\shift$ in the right-hand side is required for the
equation to be well-typed: 
\begin{equation}\label{eq:circeq-rn-last}
\small
\uf\;(\otL\;f) \circeq \JJC{()}\;(\uf \;f)
\qquad\qquad
(\text{for } f : \stseq{J}{J',\Gamma}{C})
\end{equation}

The two cut rules in (\ref{eq:cut}) are admissible in the right-normal
sequent calculus. In this case, they need to be defined by mutual
induction with another cut rule
\[
\small
\infer[\ccutj]{\stseq S {\Delta_0, A',\Gamma, \Delta_1} C}{
  \stseq {A'} {\Gamma} A
  &
  \stseq S {\Delta_0, A, \Delta_1} C
}
\]

In the fully skew sequent calculus of
Section~\ref{sec:seqcalcskewmon}, the rule $\ccutj$ is definable by
first applying $\uf$ to the first premise and then using $\ccut$.  In
the right-normal case, we have to define it simultaneously with
$\scut$ and $\ccut$ because of the added cases for the added primitive
rules. The definition of $\ccutj$ relies on a lemma, stating that, if a
sequent $\stseq A {\Gamma} J$ is derivable, with $J$ a closed formula,
then both $A$ and all the formulae in $\Gamma$ are also
closed.

The interpretation $\sound$ of fully skew sequent calculus derivations as
categorical calculus derivations extends to the right-normal
case. Given $f : \stseq{S} {\Gamma_0,\Gamma_1}{C}$, 
define $\sound\;(\IC {\Gamma_0} \;f)$ as:
\[
\small
\infer={\asem {S} {\Gamma_0,\I,\Gamma_1} \tto C}{
  \infer[\dcomp]{\asem{\asem{S}{\Gamma_0} \ot \I}{\Gamma_1} \tto C}{
    \infer[\morsem{\Gamma_1}]{\asem{\asem{S}{\Gamma_0} \ot \I}{\Gamma_1} \tto \asem{\asem{S}{\Gamma_0}}{\Gamma_1}}{
      \infer[\rhoinv]{\asem{S} {\Gamma_0} \ot \I \tto \asem{S} {\Gamma_0}}{}
    }
    &
    \infer={\asem{\asem{S}{\Gamma_0}}{\Gamma_1} \tto C}{
      \asem{S}{\Gamma_0,\Gamma_1} \overset{\sound\,f}{\tto} C
    }
  }
}
\]
The double-line rules correspond to applications of the provable
equality $\asem {S} {\Gamma,\Delta} =
\asem{\asem{S}{\Gamma}}{\Delta}$, while $\morsem{~}$ is
the inference rule introduced in (\ref{eq:morsem}).

Given $f : \stseq{S} {\Gamma_0,J,J',\Gamma_1}{C}$, 
define $\sound\;(\JJC {\Gamma_0} \;f)$ as:
\begin{equation}\label{eq:sound-otC}
\small
\infer={\asem {S} {\Gamma_0,J \ot J',\Gamma_1} \tto C}{
  \infer[\dcomp]{\asem{\asem{S}{\Gamma_0} \ot (J \ot J')}{\Gamma_1} \tto C}{
    \infer[\morsem{\Gamma_1}]{\asem{\asem{S}{\Gamma_0} \ot (J \ot J')}{\Gamma_1} \tto \asem{(\asem{S}{\Gamma_0} \ot J) \ot J'}{\Gamma_1}}{
      \infer[\alinvc]{\asem{S} {\Gamma_0} \ot (J \ot J') \tto (\asem{S} {\Gamma_0} \ot J) \ot J'}{}
    }
    &
    \infer={\asem{(\asem{S}{\Gamma_0} \ot J) \ot J'}{\Gamma_1} \tto C}{
      \asem{S}{\Gamma_0,J,J',\Gamma_1} \overset{\sound\,f}{\tto} C
    }
  }
}
\end{equation}
The derivation $\alinvc : A \ot (J \ot J') \tto (A \ot J) \ot J'$ is
the inverse of a restricted form of the associator $\al$ in which the
second and third formula are closed. It is defined and shown to invert
$\al$ by induction on $J'$ (distinguishing the two cases of $J'$ being
$\I$ and the tensor of two closed formulae).

Let us mention that, instead of $\IC{}$ and $\JJC{}$, one could of course
choose to work with one ``big-step'' inference rule 
\[
\small
\infer[\mathsf{I}_{\Gamma_0}]{\stseq{S} {\Gamma_0,J,\Gamma_1}{C}}{
  \stseq{S}{\Gamma_0,\Gamma_1}{C}
}
\]
but we prefer the ``small-step'' $\IC{}$ and $\JJC{}$ especially
because, in the situation of simultaneous right- and
associative-normality, $\JJC{}$ is subsumed by
the $\otC{}$ rule that
we will introduce in the next section.



Again, the congruence relation $\circeq$ read as a term rewrite system
is weakly confluent and strongly normalizing, and normal-form
derivations in the sequent calculus can be described as derivations in
a focused subsystem. Sequents in the focused calculus are
annotated by three possible phase annotations. Phases $\mathsf{L}$ and
$\mathsf{R}$ are as in the fully skew focused calculus
(\ref{eq:focusseqcalc}). In the new phase $\mathsf{C}$, sequents are
of the form $\stseqC S {\Omega} {\Gamma} C$, where the context is
split in two parts: an \emph{anteroom} $\Omega$ and a passive context
$\Gamma$. In phase $\mathsf{C}$, each formula $D$ in the anteroom is inspected,
starting from the right end of the anteroom. In case $D$ is the unit,
then it is removed from the anteroom.
If $D$ is a tensor $J \ot J'$, with $J$ and $J'$ closed formulae, 
then it is decomposed and $J'$ is inspected next.
Otherwise, $D$ is moved to the left
end of the passive context. 

\begin{equation}\label{eq:focusseqcalc-rn}
  \small
\begin{array}{c}
\hspace*{-4mm}
  \infer[\IC{}]{\stseqC S {\Omega, \I} {\Gamma} C}{
    \stseqC S {\Omega} {\Gamma} C
  }
  \quad
  \infer[\JJC{}]{\stseqC S {\Omega, J \otimes J'} {\Gamma} C}{
    \stseqC S {\Omega, J, J'} {\Gamma} C
  }
  \quad 
  \infer[\act]{\stseqC S {\Omega,D} {\Gamma} C}{
    \stseqC S {\Omega} {D,\Gamma} C & D \not= J 
  }
  \quad
  \infer[\mswLC]{\stseqC S {~} {\Gamma} C}{
    \stseqL S {\Gamma} C
  }
  \\[6pt]
  \infer[\uf]{\stseqL{\n}{A, \Gamma} C}{
    \stseqL A {\Gamma} C
  }
  \quad\;\;
  \infer[\IL]{\stseqL{\I}{\Gamma}{C}}{
    \stseqL{\n}{\Gamma} C
  }
  \quad\;\;
  \infer[\otL]{\stseqL{A \ot B}{\Gamma} C}{
    \stseqC A B {\Gamma} C
  }
  \quad\;\;
  \infer[\mswRL]{\stseqL T {\Gamma} C}{
    \stseqR T {\Gamma} C
  }
  \\[6pt]
  \infer[\ax]{\stseqR X {~} X}{
  }  
  \qquad
  \infer[\IR]{\stseqR{\n}{~} {\I}}{
  }
  \qquad
  \infer[\otR]{\stseqR T {\Gamma, \Delta} {A \otimes B}}{
    \stseqR T {\Gamma} A
    &
    \stseqL{\n} {\Delta} B
  } 
\end{array}
\end{equation}

All $\mathsf{C}$-phase, $\mathsf{L}$-phase and $\mathsf{R}$-phase
sequents in a derivation of a sequent $\stseqC S {\Omega} {~} C$ have
the (passive) context free of closed formulae.

By dropping the phase annotations (also turning $\dottedmid$ into a
comma), we can define three functions $\emb_{\mathsf{C}}$,
$\emb_{\mathsf{L}}$ and $\emb_{\mathsf{R}}$ embedding
right-normal focused calculus derivations into the
unfocused calculus. We can also define a normalization function
$\focus : \stseq S {\Omega} C \to \stseqC S {\Omega} {~} C$, which
identifies $\circeq$-related derivations. The functions $\focus$ and
$\emb_\mathsf{C}$ (restricted to sequents whose passive context is empty) establish a bijection between the right-normal 
sequent calculus and its focused subsystem.

The central design element of this focused calculus, the anteroom,
together with the associating organization of the phases, is due to
Chaudhuri and Pfenning~\cite{CP:fociml}.

In the right-normal focused calculus, the type (i) non-determinism in
phase $\mathsf{L}$ between the $\uf$ and $\msw$ rules is
present. But the type (ii) non-determinism in phase $\mathsf{R}$ 
concerning the split of the context at $\otR$ is inessential. Since
the context cannot contain any closed formulae, at most one of the
splits of the context into $\Gamma$, $\Delta$ can lead to a complete
derivation.

In the free right-normal skew monoidal category, we have
$\id \doteq \rho \ot \lambda \comp \alpha : (X \ot \I)
\ot Y \tto (X \ot \I) \ot Y$:
\[
\id
\doteq \rho \ot \id \comp \rhoinv \ot \id
\doteq \rho \ot \id \comp (\id \ot \lam \comp \al \comp \rho \ot \id) \comp \rhoinv \ot \id
\doteq (\rho \ot \lambda) \comp \alpha
\]
There is accordingly a single focused derivation of the sequent 
$\stseqC{(X \ot \I) \ot Y}{~}{~}{(X \ot \I) \ot Y}$.
Compare this with the two distinct derivations of the sequent in the
fully skew focused calculus, displayed in~(\ref{eq:nondet2}). In the right-normal focused calculus, the unit is removed from the antecedent (more precisely, from the anteroom) with an application of the $\IC{}$ rule, so the $\otR$ does
not have to choose in which premise to send it.
\[
\small 
\infer[\mswLC]{\stseqC {(X \ot \I) \ot Y} {~}{~} {(X \ot \I) \ot Y}}{
  \infer[\otL]{\stseqL {(X \ot \I) \ot Y} {~} {(X \ot \I) \ot Y}}{
    \infer[\act]{\stseqC {X \ot \I}{Y} {~} {(X \ot \I) \ot Y}}{
      \infer[\mswLC]{\stseqC {X \ot \I}{~} {Y} {(X \ot \I) \ot Y}}{
        \infer[\otL]{\stseqL {X \ot \I} {Y} {(X \ot \I) \ot Y}}{
          \infer[\IC{}]{\stseqC {X}{\I} {Y} {(X \ot \I) \ot Y}}{
            \infer[\mswLC]{\stseqC {X}{~} {Y} {(X \ot \I) \ot Y}}{
              \infer[\mswRL]{\stseqL {X} {Y} {(X \ot \I) \ot Y}}{
                \infer[\highlight{\otR}]{\stseqR {X} {Y} {(X \ot \I) \ot Y}}{
                  \infer[\otR]{\stseqR {X} {~} {X \ot \I}}{
                    \infer[\ax]{\stseqR {X} {~} {X}}{}
                    &
                    \infer[\mswRL]{\stseqL{\n}{~}{\I}}{
                      \infer[\IR]{\stseqR{\n}{~}{\I}}{}
                    }
                  }
                  &
                  \infer[\uf]{\stseqL {\n} {Y} {Y}}{
                    \infer[\mswRL]{\stseqL {Y} {~} {Y}}{
                      \infer[\ax]{\stseqR {Y} {~} {Y}}{}
                    }
                  }
                }
              }
            }
          }
        }
      }
    }
  }
}
\]

\subsection{Associative-Normality}
\label{sec:assoc}


The free associative-normal skew monoidal category $\FskAN(\At)$ on a set
$\At$ is obtained by extending the grammar of derivations in the
fully skew categorical calculus (\ref{eq:fsk}) with a new inference rule:
\[
\small
\infer[\alinv]{A \ot (B \ot C) \tto (A \ot B) \ot C}{}
\]
and extending the equivalence of derivations (\ref{eq:doteq}) with two
new equations: $\al \comp \alinv \doteq \id$ and $\alinv \comp \al \doteq \id$.

The associative-normal sequent calculus is
obtained by adding to (\ref{eq:seqcalc}) a new logical left rule for
the tensor, relaxing the condition for decomposing a formula $A \ot B$
in the antecedent:
\[
\small
\infer[\otLcxt {\Gamma_0}]{\stseq{S} {\Gamma_0,A\ot B,\Gamma_1}{C}}{
  \stseq{S}{\Gamma_0,A,B,\Gamma_1}{C}
}
\]
Including the rule $\otLcxt{}$
in the calculus makes it possible to derive the sequent corresponding to
$\alinv : A \ot (B \ot C) \tto (A \ot B) \ot C$:
\[
\small
\infer[\otL]{\stseq{A \ot (B \ot C)}{~}{(A \ot B) \ot C}}{
  \infer[\otC {()}]{\stseq{A} {B \ot C}{(A \ot B) \ot C}}{
    \infer[\otR]{\stseq{A}{B,C}{(A \ot B) \ot C}}{
      \infer[\otR]{\stseq{A}{B}{A \ot B}}{
        \infer[\ax]{\stseq{A}{~}{A}}{}
        &
        \infer[\uf]{\stseq{\n}{B}{B}}{
          \infer[\ax]{\stseq{B}{~}{B}}{}
        }
      }
      &
      \infer[\uf]{\stseq{\n}{C}{C}}{
        \infer[\ax]{\stseq{C}{~}{C}}{}
      }
    }
  }
}
\]

We do not include here all the new equations that need to be added to
the ones in (\ref{eq:circeq}) as generators of the least congruence
$\circeq$. They are obtained from the equalities in
(\ref{eq:circeq-rn}) (except for the 1st and 2rd) by
replacing $\IC{}$ and $\JJC{}$ with $\otLcxt{}$ and the equation
(\ref{eq:circeq-rn-last}) by replacing $\JJC{}$ with $\otLcxt{}$.

The cut rules in (\ref{eq:cut}) are also admissible in the
associative-normal sequent calculus. In this case, they need to be
defined by mutual induction with a third cut rule:
\[
\small
\infer[\ccuts]{\stseq S {\Delta_0, \ssem{S'} ,\Gamma, \Delta_1} C}{
  \stseq {S'} {\Gamma} A
  &
  \stseq S {\Delta_0, A, \Delta_1} C
}
\]
The definition of $\ccuts$ 
relies on an additional admissible rule that is a restricted form of the
$\IC{}$ rule in the right-normal sequent calculus: a unit in the
context can be removed, provided that the part of the context on its
right is non-empty.
\[
\small
\infer[\ICn{\Gamma_0}]{\stseq{S} {\Gamma_0,\I,B,\Gamma_1}{C}}{
  \stseq{S}{\Gamma_0,B,\Gamma_1}{C}
}
\]

The case of the function $\sound$ for the new inference rule $\otC{}$
is defined as for the rule $\JJC{}$ in the right-normal sequent calculus,
given in (\ref{eq:sound-otC}), with the closed formulae $J$ and $J'$
replaced by arbitrary formulae, and the application $\alinvc$ replaced by $\alinv$.


We need not prove soundness for the rule $\ICn{}$ since it is not
primitive. But it is motivated by the presence in the fully skew
categorical calculus already of a derivation
$(\id \ot \lam) \comp \alpha : (A \ot \I) \ot B \tto A \ot B$ that, 
by (m2), post-inverts $\rho \ot \id$.

An equivalent focused subsystem is obtained similarly to the
right-normal case. It is in fact the same as the focused calculus in
(\ref{eq:focusseqcalc-rn}) with the rule $\IC{}$ removed, the rule
$\JJC{}$ replaced by $\otLcxt{}$ and the rule $\act$ modified
accordingly. If the rightmost formula $D$ in
the anteroom is a tensor $A \ot B$, then it is decomposed and $B$ is
inspected next. Otherwise, $D$ is moved to the left end of the passive
context.
\[
\small
\infer[\otLcxt{}]{\stseqC S {\Omega,A \ot B} {\Gamma} C}{
    \stseqC S {\Omega,A,B} {\Gamma} C
}
\qquad
\infer[\act]{\stseqC S {\Omega,D} {\Gamma} C}{
  \stseqC S {\Omega} {D,\Gamma} C & D \not= A \ot B
}
\]

All $\mathsf{C}$-phase, $\mathsf{L}$-phase and $\mathsf{R}$-phase
sequents in a derivation of a sequent $\stseqC S {\Omega} {~} C$ have
the (passive) context free of formulae $A \ot B$.

Similarly to the right-normal case discussed in Section~\ref{sec:right}, the
associative-normal sequent calculus can be proved
equivalent to its focused subsystem by means of two functions
$\emb_{\mathsf{C}}$ and $\focus$.

Lack and Street \cite{LS:triosm} observed that the free associative-normal skew monoidal category on one generator $\FskAN{(1)}$ is isomorphic to $\Delta_\bot$, the category of finite non-empty ordinals and first-element-and-order-preserving functions, and their main coherence theorem states that the canonical functor $\Fsk{(1)} \to \Delta_\bot$ is faithful.
This seems to imply that the embedding of the fully skew sequent calculus into the associative-normal sequent calculus is faithful, but we leave a direct proof of this fact to future work.
\nz{p.s. with respect to the name of the functor: hmm. I see your point, though Lack and Street call it a ``forgetful'' functor!}

\subsection{Multiple Normality Conditions}

The additional inference rules and equations for the three normality
aspects can be enabled simultaneously, to yield, e.g., a sequent
calculus for the free simultaneous left-normal and right-normal skew monoidal
category.

The following rules define a single focused sequent calculus that can
handle any combination of the three normality aspects. It is
parameterized in three flags $ln$, $rn$ and $an$ for left-, right-
resp.\ associative normality. (Recall that $J$, $J'$ are
metavariables for closed formulae.)
\[
\small
\begin{array}{c}
  \infer[\IC{}]{\stseqC S {\Omega,\I} {\Gamma} C}{
    \stseqC S {\Omega} {\Gamma} C
    &
    rn
  }
  \qquad
  \infer[\otLcxt{}]{\stseqC S {\Omega,A \ot B} {\Gamma} C}{
    \stseqC S {\Omega,A,B} {\Gamma} C
    &
    (rn \wedge A \ot B = J \ot J') \vee an
  }
  \\[6pt]
  \infer[\act]{\stseqC S {\Omega,D} {\Gamma} C}{
    \stseqC S {\Omega} {D,\Gamma} C
    &
    rn \to D \not= J
    &
    an \to D \not= A \ot B
  }
  \qquad
  \infer[\mswLC]{\stseqC S {~} {\Gamma} C}{
    \stseqL S {\Gamma} C
  }
  \\[6pt]
  \infer[\uf]{\stseqL{\n}{A, \Gamma} C}{
    \stseqL A {\Gamma} C
  }
  \quad
  \infer[\IL]{\stseqL{\I}{\Gamma}{C}}{
    \stseqL{\n}{\Gamma} C
  }
  \quad
  \infer[\otL]{\stseqL{A \ot B}{\Gamma} C}{
    \stseqC A B {\Gamma} C
  }
  \quad 
  \infer[\mswRL]{\stseqL T {\Gamma} C}{
    \stseqR T {\Gamma} C
    &
   ln \wedge T = \n \to \Gamma = ()
  }
  \\[6pt]
  \infer[\ax]{\stseqR X {~} X}{
  }  
  \quad
  \infer[\IR]{\stseqR{\n}{~} {\I}}{
  }
  \quad  
  \infer[\otR]{\stseqR T {\Gamma, \Delta} {A \otimes B}}{
    \stseqR T {\Gamma} A
    &
    \stseqL{\n} {\Delta} B
  }
  \quad
  \infer[\otRem]{\stseqR {X} {\Delta} {A \otimes B}}{
    \stseqR {\n}{~} A
    &
    \stseqR{X} {\Delta} B
    & ln
  }
\end{array}
\]

In the case of simultaneous left- and right-normality (i.e.,
$ln \wedge rn$), the non-determinism of type (i) in phase $\mathsf{L}$
is not present and the non-determinism of type (ii) in phase
$\mathsf{R}$ is inessential as there cannot be any closed
formulae in the context. Consequently, any sequent can have at most
one focused derivation: the free simultaneously left- and right-normal
skew-monoidal category is thin.

Laplaza \cite{Lap:cohani} proved that the free skew semigroup category is
thin (and this was reproved by Zeilberger~\cite{Zei:seqcsa}). We have now seen that this remains true also when a 
both left-normal and right-normal unit is freely added.

\section{Conclusions and Future Work}

We showed that, similarly to the free skew monoidal category and the free
monoidal category, the free skew monoidal categories
of different degrees of partial normality can be described as sequent
calculi. These sequent calculi define logics weaker than the
multiplicative fragment of the intuitionistic non-commutative linear
logic. They enjoy cut elimination and they also admit focusing,
a deductive description of a root-first proof search strategy that finds
exactly one representative from each equivalence class of derivations.

We intend to continue this study by broadening its scope to fully skew
and partially normal closed and monoidal closed categories and also
prounital-closed categories (where the unit is present in a
non-represented way). Skew closed categories (the skew variant of closed categories of Eilenberg and Kelly \cite{EK:cloc}) were first considered by
Street \cite{Str:skecc}, prounital-closed categories by Shulman
\cite[Rev.~49]{ncatlab:cloc}. Zeilberger \cite{Zei:theltf} used a thin variant of skew closed categories, which he called imploids, in his study of the relation between typing of linear lambda terms and flows on 3-valent graphs.
In the recent work \cite{UVZ:eilkr}, we
dissected the Eilenberg-Kelly theorem about adjoint monoidal and
closed structures on a category, revisited by Street \cite{Str:skecc}
for the skew situation, to establish it in the general partially
normal case. Some surprising phenomena occur around skew closed
categories, e.g., the free skew closed category on a set is left-normal, but
this is lost when the tensor is added. We also want to study the proof
theory of skew braided monoidal categories, as recently introduced by
Bourke and Lack \cite{BL:brasmc}.

We have in this paper explained how the free skew monoidal category
of each possible degree of partial normality can be described as a sequent calculus (a ``logic''). This
correspondence extends to non-free partially normal
monoidal categories. But in this case, rather than inductively
defining the maps and their equality, the inference rules and
equations of the sequent calculus merely impose closure conditions on
some homset predicate and some equality relation given upfront. One could
compare a non-free category to a ``theory'' (a set of judgments closed under some inference rules) as opposed to a
``logic'' (the least set of judgments closed under them, i.e., the set of derivable judgments). Cut elimination (in the sense that a set of judgments closed under the inference rules adopted minus cut would necessarily be closed also under cut) cannot be expected, neither can focusing.
\tu{How about something like this? That's what I meant. Like a pure logic opposed to logic plus some ``extralogical'' truths.}
\nz{The two previous sentences somehow rub me the wrong way -- I don't see ``axiomatic theory'' and ``logic'' in real opposition, and I'm not sure what it would mean for cut elimination and focusing to \emph{fail} in a non-free category.}
Bourke and Lack \cite{BL:multi} showed that skew monoidal categories
are equivalent to representable skew multicategories, a weakening of
representable multicategories \cite{Her:repm}. In our previous work
\cite{UVZ:seqcsm}, we showed that the map constructors and equations
of a (nullary-binary) representable skew multicategory are very close
to and mutually definable with those of the sequent calculus for the
corresponding skew monoidal category (viewed as a deductive calculus,
the representable skew multicategory uses exactly the same sequent forms, 
but has
the basic inference rules and equations chosen differently). We expect
that partially normal skew monoidal categories can be
analyzed in similar terms. Specifically, we hope that the correct
variations of representable skew multicategories can be
systematically derived in the framework of (op)fibrations of
multicategories \cite{Her:fibam}, adapted for skew multicategories.

\subsubsection*{Acknowledgments}

T.U.\ was supported by the Icelandic Research Fund grant no.~196323-052
and the Estonian Ministry of Education and Research institutional
research grant no.~IUT33-13. N.V.\ was supported by the ESF funded
Estonian IT Academy research measure (project 2014-2020.4.05.19-0001).





\newcommand{\doi}[1]{\href{https://doi.org/#1}{doi: #1}}

\end{document}